\documentclass[prl,aps,twocolumn,floatfix]{revtex4}
\usepackage{graphicx}

\begin{document}

\title{Effect of disorder on the conductance of a Cu atomic point contact}
\author{D. J. Bakker} \thanks{Present address: Department of Applied Physics and DIMES,
Delft University of Technology, Lorentzweg 1, 2628 CJ Delft, The
Netherlands}
\author{Y. Noat} \thanks{Present address: Groupe de Physiques des
Solides, Campus Jussieu tour 23, 2 Place Jussieu, 75251 Paris
cedex 05, France}
\author{A. I. Yanson} \thanks{Present address: Dept. of Physics, 510
Clark Hall, Cornell University, Ithaca, NY 14853}
\author{J. M. van Ruitenbeek}
\affiliation{Kamerlingh Onnes Laboratorium, Universiteit Leiden,
Postbus 9504, 2300 RA Leiden, The Netherlands}

\date{\today}

\begin{abstract}
We present a systematic study of the effect of the disorder in
copper point contacts. We show that peaks in the conductance
histogram of copper point contacts shift upon addition of nickel
impurities. The shift increases initially linearly with the nickel
concentration, thus confirming that it is due to disorder in the
nanowire, in accordance with predictions. In general, this shift
is modelled as a resistance $R_s$ which is placed in series with
the contact resistance $R_c$. However, we obtain different $R_s$
values for the two peaks in the histogram, $R_s$ being larger for
the peak at higher conductance.
\end{abstract}

\maketitle

\section{Introduction}
Quantum point contacts (QPC's) are constrictions in a conducting
material whose size is of the order of the Fermi wavelength of the
electrons. The first quantum point contacts were fabricated in a
2-dimensional electron gas (2DEG) in a semiconductor
heterostructure \cite{PRL_vanwees88,JPC_Wharam88}. Their
conductance exhibits a series of steps as a function of
constriction size. The plateaus in between these steps are
positioned at integer multiples of \mbox{$G_0 = 2e^2/h$ $\approx$
(12.9 k$\Omega$)$^{-1}$}, the quantum of conductance. This
phenomenon is explained as a consequence of the quantization of
the electronic energy levels, resulting from the confinement of
the electrons in the direction perpendicular to the current. Signs
of quantized behavior were also seen in the conductance of quantum
point contacts made from metals
\cite{PRB_Krans93,PRL_Pascual93,PRL_Olesen94}. When a thin
metallic wire is slowly broken, its conductance as a function of
contact size shows smooth plateaus, alternated by steps of the
order of $G_0$. In this case, however, the plateaus are not
necessarily horizontal or positioned at integer multiples of
$G_0$. Moreover, when the two wire ends are pressed together and
broken again, the precise shape of the trace differs from trace to
trace.  In order to obtain fully reproducible results, a histogram
is often created by projecting each conductance trace onto the
conductance axis and adding the contribution of several thousands
of traces. For many metals, the histograms show well-defined peaks
which represent preferred conductance values.

The physical origin of these peaks has been debated heavily
\cite{PRL_Yanson97}, especially the question whether they are
indeed related to the quantization of electronic energy levels or
rather to the discrete and  abrupt changes in the cross-sectional
area of the nanowires, which are caused by the fact that the
contact must change its size  by at least one, discrete, atom. It
was shown both experimentally, by force measurements
\cite{PRL_Rubio96}, and theoretically, by numerical simulations
\cite{PRL_Sutton93}, that steps are correlated with atomic
rearrangements of the contact. Thus, plateaus in the conductance
traces correspond to stable atomic configurations. Apart from its
geometry, the conductance of such a configuration is determined by
the conductance of individual atoms of the metal under
investigation. It was shown \cite{Nature_Scheer98} that the
conductance of a one-atom point contact is governed by the nature
of its valence orbitals. For monovalent s-metals such as Na, Cu or
Au, a one-atom contact has a conductance of nearly 1 $G_0$. On the
other hand, transition metals such as Ni, Pd or Pt, which have
d-valence orbitals, have a one-atom conductance of the order of
1.6 $G_0$.

In experiments on both semiconductor QPC's \cite{PRL_vanwees88}
and QPC's made of monovalent s-metals it has been observed that
the quantization of the conductance is never perfect: the peaks in
the histograms are shifted below quantized values $mG_0$ ($m$ = 1,
2, 3 $\ldots$). In metallic point contacts, such a shift has been
observed in gold \cite{PRB_Costa-Kramer97, PRB_Hansen97}, sodium
\cite{Nature_Krans95}, silver \cite{PRB_Hansen97} and copper
\cite{PRB_Hansen97}. Two possible origins of the shift have been
discussed in the literature. It has been attributed to disorder
either in the leads to the contact \cite{PRL_vanwees88}, or in the
contact itself \cite{PRL_Garcia97,PRB_Brandbyge97,PRB_Burki99}. In
all cases, a model was adopted in which the shift is accounted for
by an `extra' resistance $R_s$ that is placed in series with the
true contact resistance $R_c$. $G_c \equiv 1/R_c$ is assumed to be
an exact integer multiple of $G_0$, and it is then straightforward
to determine $R_s$ from the histograms. Typical experimental
values are 400 $\Omega$ for a 2DEG \cite{PRL_vanwees88}, 100
$\Omega$ for sodium \cite{Nature_Krans95},  100-500 $\Omega$ for
gold \cite{PRB_Costa-Kramer97,PRB_Hansen97}, 500 $\Omega$ for
silver \cite{PRB_Hansen97} and 700 $\Omega$ for copper
\cite{PRB_Hansen97}.

In all theoretical studies into disorder in QPC's, adding disorder
resulted in a downshift and smearing of the peaks in the
conductance histogram. In the work of Garc\'\i a-Mochales
\textit{et al.} \cite{PRL_GarciaMochales97}, for example, the
nanowire was simulated by a tight binding model. The disorder is
introduced by random on-site energies, while the overlap between
atomic orbitals is described by a hopping term between sites. The
conductance was calculated as a function of the Fermi energy,
while the geometry was kept constant. A histogram was then
constructed, which showed peaks that were indeed shifted. The
series resistance (calculated by fitting the best resistance value
centering simultaneously all peak maxima at the corresponding
integer values of $G_0$) were found to range from about 10
$\Omega$ for small disorder to 1000 $\Omega$ for large disorder.

B\"urki \textit{et al.} \cite{PRB_Burki99} used a 2D-free electron
model, with a potential term in the Hamiltonian describing
randomly distributed impurities. The neck of the constriction in
the nanowire was elongated, while keeping the Fermi energy
constant. The histogram exhibited downshifted peaks that could be
positioned at integer multiples of $G_0$ by subtracting a series
resistance of 525 $\Omega$.

Thus, while several authors have studied disorder in QPC's from a
theoretical point of view, very little experimental study was
devoted to it. Ludoph \textit{et al.} \cite{bas} studied
conductance fluctuations as a function of bias voltage in
atomic-size contacts for pure metals. A consistent explanation was
obtained in terms of scattering of the electrons on defects near
the contact and interference of various electron paths with the
reflected wave at the contact itself. The values obtained for the
elastic scattering length, $l_e$, were used to estimate an
effective series resistance that would result from this
scattering, and this resistance was found to be in agreement with
the observed shifts in the conductance peaks near 1 and 3 $G_0$
for Cu, Ag and Au. On the other hand, the shift of the second peak
was much larger and this led Ludoph \textit{et al.} to propose
that the origin of this peak may be a preferred atomic
configuration, rather than conductance quantization.  Li
\textit{et al.} \cite{PRB_Li98} studied the influence of the
presence of adsorbate molecules such as nitrogen or ethanol on the
conductance of a gold nanowire. Plateaus are still observed in
individual conductance traces, but their positions scatter more,
resulting in a smearing of peaks in the histogram, which depends
on the strength of the adsorbate used. This effect is interpreted
as a result of the scattering of conduction electrons by the
adsorbate molecules. It is noted that the first peak is less
affected than the others.

The work presented here is an experimental study of the effect on
conductance histograms of various levels of disorder. For this
purpose, we have measured the histograms of copper-nickel alloys
for various nickel concentrations, and compared them to the
histogram of pure copper. The disorder in the point contact is
increases with the nickel concentration.

\section{Experimental method}

Our samples are wires made of random alloys of copper and nickel.
Two reasons have motivated the choice for this alloy: Firstly, its
phase diagram is particularly simple, copper and nickel being
completely miscible in all concentrations. Secondly, the
histograms of copper and nickel are very different, as will be
shown below. This makes it easy to recognize whether increasing
the nickel concentration changes the histogram in a qualitative
way, which would make it impossible to see the nickel atoms as
impurities in the copper lattice, or whether it induces only
quantitative changes, e.g. a shift of the peaks in the copper
histogram.

The nickel concentrations used are 1, 10, 25, 35, 45 and 50 atomic
\%. The required amounts of copper and nickel were melted at 1500
$^{\circ}$C. The alloys were pulled into wires with a diameter of
about 200 $\mu$m, which were subsequently annealed for 24 hours at
900$^{\circ}$C.
\begin{figure}
\includegraphics[width = .5 \textwidth]{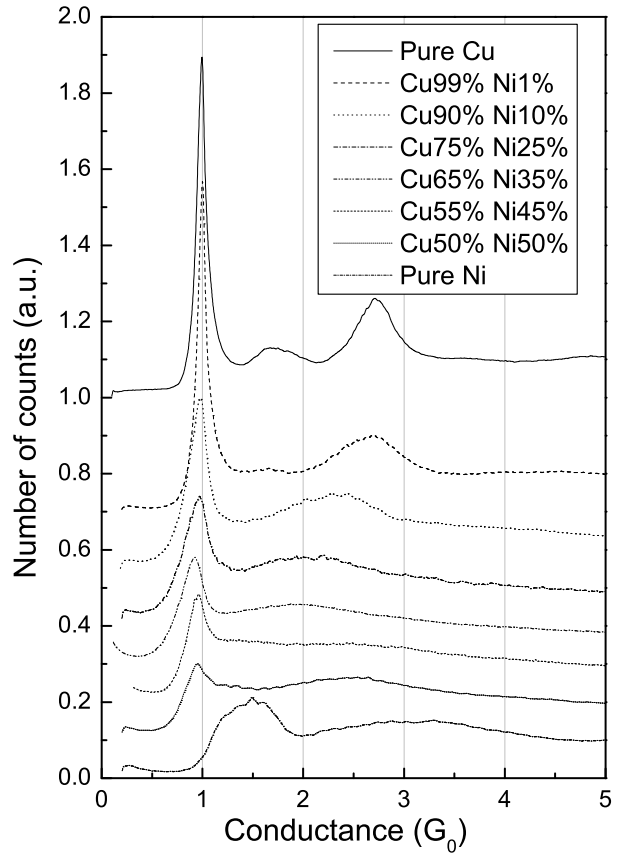}
\caption{Averaged conductance histograms at 10 mV bias. From top
to bottom: pure copper, copper alloys containing 1, 20, 25, 35, 45
and 50\% nickel, and pure nickel. All histograms have been
normalized to the area below the curve and were shifted along the
$y$-axis for clarity.} \label{comparison}
\end{figure}
Metallic contacts of atomic size were then obtained by using the
so called Mechanically Controllable Break Junction (MCBJ)
technique. A piece of wire is notched in the middle and glued on a
bendable phosphor-bronze substrate by two drops of epoxy. The
sample is insulated from the substrate by a layer of kapton foil,
and four measurement wires are connected to it with a conducting
glue. The sample is first cooled down to 4.2 K and then broken in
vacuum ($10^{-5}$ mbar at room temperature) by bending the
substrate. The bending is achieved by pressing a piezo element
against the substrate by means of a differential screw. The high
vacuum provided by the low temperature ensures atomically clean
contacts after rupture. The two electrodes are then brought back
into contact and their separation is controlled using the piezo
element. In this way, the displacement between the two electrodes
can be controlled with a precision of the order of 0.01 \AA. The
wire is broken and brought back into contact repeatedly by
applying a sawtooth voltage to the piezo element. During the
breaking stages current and voltage over the sample are measured.
An histogram is build from approximately 3000 individual
conductance traces. To ensure reproducibility, for each alloy at
least two different samples were measured, while the contact was
regularly `renewed' by pressing the electrodes deep into each
other.

\section{Results}

In Fig. \ref{comparison}, the histograms of the pure elements Cu
and Ni and those of the copper alloys containing 1, 10, 25, 35, 45
and 50\% nickel are presented. These measurements were done using
a bias voltage of 10 mV. Topmost, the histogram of pure copper is
shown. It is obtained by averaging  ten histograms made on four
different samples of pure copper. Three peaks can be
distinguished, positioned at conductance values $G$ = 1.0 $G_0$,
1.7 $G_0$ and 2.7 $G_0$. The first peak has a higher amplitude and
is sharper than the others. Close inspection makes clear that this
first peak is actually slightly below 1 $G_0$. At the bottom of
Fig. \ref{comparison}, the histogram of pure nickel is shown. It
is very different from the copper histogram: it exhibits a broad
first peak centered around 1.6 $G_0$, and a second, even broader,
structure centered around 3.1 $G_0$.

Looking at the histograms of the alloys, it is remarkable that for
all nickel concentrations the histograms resemble the histogram of
pure copper; in particular, they have a first peak a little below
$1G_0$. Indeed, for biases lower than 200 mV and for nickel
concentrations up to 50\%, there is no sign of a "nickel" peak at
1.6 $G_0$. We give a possible interpretation of this effect in the
next section. The second peak disappears rapidly when one
increases the nickel concentration, due to shifting and
broadening. With increasing nickel concentrations, three things
can be seen to happen to the other peaks in the histogram:
broadening, decreasing in amplitude and downshifting.
\begin{figure}
\includegraphics[width = .5 \textwidth]{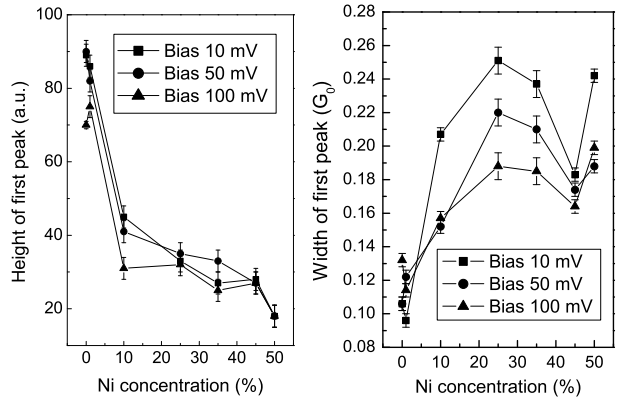}
\caption{Dependence on Ni concentration of the height and width of
the first peak in the histogram for several biases.}
\label{peakheights}
\end{figure}
\begin{figure}
\includegraphics[width = .5 \textwidth]{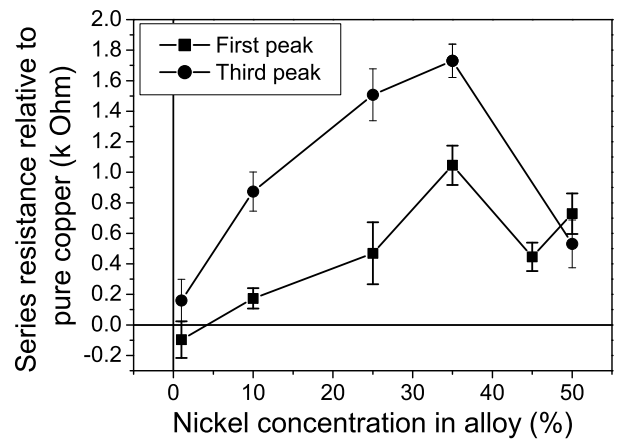}
\caption{Series resistance corresponding to the shift relative to
pure copper of the first and third peaks in the histograms for
different nickel concentrations, at a bias of 10 mV. Axes are
indicated for clarity.} \label{shiftpeaks}
\end{figure}
The width and height of the first peaks in the histogram are
plotted in the right and left panel of Fig. \ref{peakheights},
respectively. As can be seen, the amplitude of the peak decreases
monotonously with Ni concentration. This can be explained as the
result of increasing disorder. The disorder leads to slight
variations in the last-plateau conductance of different scans.
These variations cause the different plateaus to fall into
different bins on the conductance axis, thus causing less high
peaks in the histogram. The same effect gives rise to the
broadening of the peaks that is observed for Ni concentrations up
to 35\%. Above this concentration, the width decreases to a
minimum at 45\%, after which it increases again. Fig.
\ref{peakheights} also shows that the results reproduce for
different biases. The main feature observed in Fig.
\ref{comparison} is the shift towards lower conductance values of
the first and third peaks in the histogram. In Fig.
\ref{shiftpeaks}, the series resistance corresponding to this
shift is plotted as a function of the percentage of nickel in the
alloy. Note that we do not calculate the shift with respect to
integer multiples of $G_0$, but relative to the measured positions
of the corresponding peaks in the histogram of pure Cu. Results
again reproduce for other biases. The shift increases roughly
linearly with nickel concentration, up to 35\% nickel. Two
surprising features of Fig. \ref{shiftpeaks} must be noted.
Firstly, the series resistance is not the same for all the peaks
in the histogram. Instead, the resistance corresponding to the
shift of the third peak is roughly two or three times higher than
the corresponding values for the first peak. The same conclusion
can be reached when considering the resistance necessary to bring
the first and third peaks of the pure copper histogram to the
corresponding integer multiples of $G_0$, namely, 60 $\pm$ 20
$\Omega$ and 460 $\pm$ $60 \Omega$ respectively. This is in
agreement with earlier findings \cite{PRB_Hansen97}, but
contradictory to the widespread practice to bring all conductance
peaks to integer multiples of $G_0$ by subtracting a constant
series resistance. The second surprising feature of Fig.
\ref{shiftpeaks} is the decrease of the shift for Ni
concentrations higher than 35\%. This decrease corresponds to the
decrease in width that was already noticed in Fig.
\ref{peakheights}. Again the minimum is at 45\% Ni. We discuss a
possible interpretation of this effect in the next section.

\section{Discussion}

The main difference between the histograms of copper and nickel,
as shown in Fig. \ref{comparison}, is the position of the peaks.
In particular, whereas copper shows a sharp first peak nearly
exactly at 1 $G_0$, nickel has a broad first peak centered around
1.6 $G_0$. In both cases, however, it seems reasonable to assume
that the first peak in the histogram corresponds to the smallest
possible contact: only one atom bridging the gap between the
electrodes. This is supported by the fact that the conductance
invariably drops sharply to zero at the end of the plateau at
$\sim$ 1 (Cu) or $\sim$ 1.6 $G_0$ (Ni). The difference in
conductance between the different kind of atoms can be explained
by taking into account their valence orbitals. As already
mentioned in the introduction, copper has one spherically
symmetric s valence orbital, whereas the valence orbitals of
nickel are d-type. It has been shown \cite{Nature_Scheer98} that
for monovalent s-type metals there is one, almost completely
opened, conductance channel per atom, which gives rise to a 1-atom
conductance of nearly 1 $G_0$. In contrast, the five valence
orbitals of a d-metal give rise to five partially opened channels,
with a total conductance in the range of 1.5 - 2.5 $G_0$.

We noticed that the histograms for all the alloys resemble the
histogram of pure copper, in particular, the first peak is below 1
$G_0$, and the third is below 2.7 $G_0$. However, if we very
crudely model a contact of e.g. Cu$_{50}$Ni$_{50}$ as consisting
of two pyramids touching each other in the apexes, such that there
are two constricting atoms, 25\% of all contacts should end in a
Cu-Cu constriction, 25\% should end in a Ni-Ni constriction, and
50\% should have a Cu-Ni constriction. This raises the question
why we don't see the influence of the nickel atoms in the
histogram of this alloy, apart from the downshifting and smearing
of its peaks. An explanation for this phenomenon might be that the
Ni atoms do not take part in the forming of the contact. This
might be due to surface segregation, which is known to lead to a
very strong copper enrichment of the surface layers in bulk Cu-Ni
\cite{PRB_Sakurai86}, due to the lower surface energy of Cu atoms.
Calculations for small Cu-Ni clusters \cite{PRB_Tomanek83} show
the same phenomenon: all the copper atoms present segregate to the
surface of the cluster. All the atoms forming the contact in a
MCBJ can be viewed as being at the `surface', to the extent that
their coordination number is less than in the bulk. Thus there
might be a preference for Cu atoms to form the contact, which
would lead to the absence of Ni peaks in the histogram of the
alloy. However, this surface segregation takes places only at
elevated ($>$ 600 K) temperatures, while we performed our
measurements at 4 K and at low (10 mV) bias, such that local
heating of the contact is probably not sufficient for segregation
to happen \cite{helko}.  Moreover, preliminary studies by
molecular dynamics simulations \cite{jan} don't show such a
segregation between Cu and Ni atoms in the contact. Possibly,
however, the repeated contact indentation and breaking provides
enough mobility for the atoms to obtain some degree of
segregation.

Another possibility to explain the similarity between the
histograms of the alloys and that of pure copper is assuming that
the conductance of Cu-Ni constriction is the same as that of a
Cu-Cu constriction, namely, 1 $G_0$. For a contact of
Cu$_{50}$Ni$_{50}$ in the form of two pyramids, this would lead to
at least 75\% of all last-atom contacts having a conductance of
approximately 1 $G_0$. Calculations to verify this are under way.

If the above picture is correct, adding nickel atoms indeed
doesn't change the copper histogram in a qualitative way. Instead,
the nickel atoms only act as scatterers for the electrons in the
banks to the contact. One would expect the amount of disorder in
the banks to increase with the nickel concentration. In Fig.
\ref{shiftpeaks}, it is shown that the series resistance increases
approximately linearly with nickel concentration up to a nickel
concentration of 35\%. From this we conclude that it is indeed the
disorder in the nanowire, in the form of impurities, that causes
the downshift of the peaks in the histograms. This is in
accordance with the results of the calculations cited above.

\begin{figure}
\includegraphics[width = .5 \textwidth]{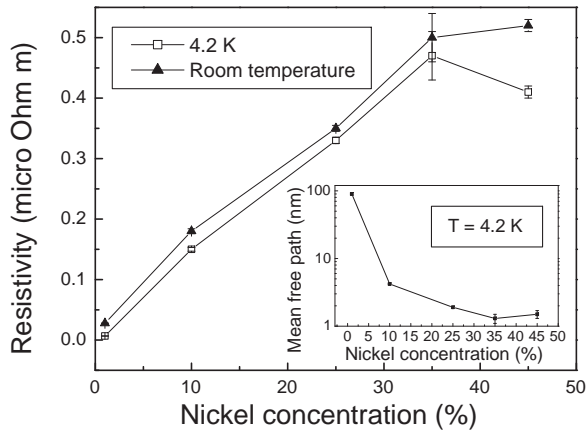}
\caption{Room temperature and low temperature (4.2 K) resistivity
of the copper-nickel alloys as function of nickel concentration.
Inset shows the bulk mean free path, calculated from a Drude
model.} \label{resistivity}
\end{figure}

However, the sudden decrease of the series resistance when going
to 45 \% nickel, accompanied by a decrease of the width of the
first peak, suggest that the amount of disorder decreases when one
increases the nickel concentration to this value. In order to
compare with the behavior of bulk copper-nickel, we have measured
the bulk resistivity of the alloys at room temperature as well as
at $T$ = 4.2 K. Results are presented in Fig. \ref{resistivity}.
It can be seen that the low-temperature resistivity initially
closely follows the RT resistivity and is linear with the nickel
concentration. Upon going to 45\% nickel, however, the
low-temperature resistivity decreases, whereas the room
temperature resistivity still increases, although less than would
be expected from a linear behavior. The decrease in the shift of
the peaks in the copper-nickel histograms therefore probably
reflects the decrease is resistivity of bulk copper-nickel. This
shows that the origin of the series resistance is probably in the
banks to the contact, and is not a (quantum mechanical) effect in
the contact itself. A likely explanation for the decrease is the
onset of ferromagnetism in the nanowire. The ferromagnetic
ordering will decrease the amount of spin-spin scattering events
by decreasing the magnetic fluctuations, even though the nickel
concentration increases. As a matter of fact, bulk copper-nickel
becomes magnetic at 4.2 K for a nickel concentration of
approximately 45\% \cite{handbook}. From Fig. \ref{shiftpeaks} it
can be seen that the resistance tends to increase again when the
nickel concentration is increased to 50\%, which may be explained
by assuming that once the alloy is ferromagnetic, the amount of
disorder in it again increases with increasing nickel
concentration. The inset in Fig. \ref{resistivity} shows the bulk
mean free paths for the electrons in several of the alloys,
calculated from the resistivities using the Drude model. An
important difference between the bulk material and the quantum
point contacts is the behavior for low nickel concentration. It is
well known that the resistivity of metals at low temperatures
depends sensitively on the purity of the sample. Thus, the mean
free path in pure copper at 4.2 K will be much larger than that in
Cu$_{99}$Ni$_1$. On the other hand, the histograms of these two
materials are nearly indistinguishable. This means that in point
contacts of both materials, the mean free path is restricted by
scattering from the surface and lattice defects related to the
geometry of the contact, rather than by impurities. When the
concentration of impurities is raised to 10\%, they start to have
an appreciable effect. Thus, the mean free path in the contact
region of a QPC of pure copper must be in between those in bulk
Cu$_{99}$Ni$_1$ and Cu$_{90}$Ni$_{10}$, i.e. between 5 and 30 nm.
This is consistent with results obtained earlier for quantum point
contacts of copper \cite{bas}.

From Fig. \ref{shiftpeaks} a second important feature can be
inferred. It can be seen clearly that the series resistance is not
the same for all the peaks in the histogram. The resistance
necessary to shift the third peak in the histograms back to its
value for pure copper is larger than for the first peak. The
difference increases with disorder. This effect is not accounted
for by any of the calculations cited above. Possibly this
difference in behavior compared to the first peak can be accounted
for by considering the fluctuations in Cu/Ni ratio for contacts of
this size, or by a non-uniform Ni concentration due to a partial
segregation of Ni atoms away from the contact area.

\section{Conclusion}

In conclusion, we have shown, firstly, that adding nickel
impurities to copper leads to an increase of the shift and
smearing of the conductance peaks. This is evidence that disorder
in the nanowires is a main source for the series resistance, in
agreement with several theoretical papers. Up to a nickel
concentration of 35\%, this shift increases roughly linearly with
the nickel concentration. Secondly, when the nickel percentage is
increased further, the shift decreases. The latter effect may be
due to a ferromagnetic transition leading to a decrease of the
resistance of the mesoscopic banks. It was shown that the bulk
resistivity of Cu-Ni alloys follows the same behavior as the
series resistance does, which is evidence that the series
resistance finds its origin in the banks to the contact rather
than in the contact itself. Thirdly, we have shown that one single
series resistance is not sufficient to shift all the peaks in the
histogram back to the respective quantized values. Instead, it is
found that the series resistance needed for the third peak is
larger than that for the first. Fourthly, our measurements show
that adding nickel impurities has remarkably little influence on
the global shape of the copper histogram, even if the nickel
concentration is raised to 50\%. To explain this, we suggest that
a Cu-Ni constriction has approximately the same conductance as a
Cu-Cu constriction, or that there is a partial segregation of Ni
atoms away from the contact.

\section{Acknowledgements}

This research has been supported by a Marie Curie fellowship of
the European Community under contract number HPMF-CT-1999-00196.
We are grateful to I.K. Yanson for fruitful discussions and to
R.W.A. Hendrikx for providing the sample materials.

\end{document}